\documentclass[%
 aip,
 amsmath,amssymb,
 reprint,%
]{revtex4-1}

\usepackage{graphicx}
\usepackage{dcolumn}
\usepackage{bm}

\usepackage[utf8]{inputenc}
\usepackage[T1]{fontenc}
\usepackage{mathptmx}
\usepackage{etoolbox}

\usepackage{dsfont}
\usepackage[caption=false]{subfig}
\renewcommand{\vec}[1]{\boldsymbol{#1}}
\newcommand{\oH}{\hat{H}}

\newcommand{\oS}{\hat{S}}
\newcommand{\ovS}{\vec{\hat{S}}}

\newcommand{\oa}{\hat{a}}
\newcommand{\oad}{\hat{a}^\dagger}
\newcommand{\ona}[1]{\hat{a}^\dagger_{#1}\hat{a}^{\phantom{\dagger}}_{#1} }
\newcommand{\ova}{\boldsymbol{\oa}}

\newcommand{\oal}{\hat{\alpha}}
\newcommand{\oald}{\hat{\alpha}^\dagger}
\newcommand{\ovaL}{\boldsymbol{\oal}}

\newcommand{\ob}{\hat{b}}
\newcommand{\obd}{\hat{b}^\dagger}
\newcommand{\onb}[1]{\hat{b}^\dagger_{#1}\hat{b}^{\phantom{\dagger}}_{#1} }

\newcommand{\obe}{\hat{\beta}}
\newcommand{\obed}{\hat{\beta}^\dagger}

\makeatletter
\def\@email#1#2{%
 \endgroup
 \patchcmd{\titleblock@produce}
  {\frontmatter@RRAPformat}
  {\frontmatter@RRAPformat{\produce@RRAP{*#1\href{mailto:#2}{#2}}}\frontmatter@RRAPformat}
  {}{}
}%
\makeatother
\begin{document}

\preprint{AddPreprint/here}

\title{Dipole-dipole-interaction-induced entanglement between two-dimensional ferromagnets}
\author{D. Wuhrer}
 \affiliation{Fachbereich  Physik,  Universit\"at  Konstanz,  D-78457  Konstanz,  Germany}%
 \affiliation{Authors to whom correspondence should be addressed: dennis.wuhrer@uni-konstanz.de and wolfgang.belzig@uni-konstanz.de}

\author{N. Rohling}%
 \affiliation{Fachbereich  Physik,  Universit\"at  Konstanz,  D-78457  Konstanz,  Germany}%

\author{W. Belzig}
 \affiliation{Fachbereich  Physik,  Universit\"at  Konstanz,  D-78457  Konstanz,  Germany}%
  \affiliation{Authors to whom correspondence should be addressed: dennis.wuhrer@uni-konstanz.de and wolfgang.belzig@uni-konstanz.de}

\date{\today}

\begin{abstract}
    We investigate the viability of dipole-dipole interaction as a means of entangling two distant ferromagnets. To this end we make use of the Bogoliubov transformation as a symplectic transformation. We show that the coupling of the uniform magnon modes can be expressed using four squeezing parameters which we interpret in terms of hybridization, one-mode and two-mode squeezing. We utilize the expansion in terms of the squeezing parameters to obtain an analytic formula for the entanglement in the magnon ground state using the logarithmic negativity as entanglement measure. Our investigation predicts that for infinitely large two-dimensional ferromagnets, the dipole-dipole interaction does not lead to significant long-range entanglement. However, in the case of finite ferromagnets, finite entanglement can be expected. 
\end{abstract}

\maketitle

Magnetic materials exhibit a wide range of magnetic structures \cite{Spin_Waves_Dyson, AFM_Neel, SkyrmionsBogdanov, SpinSpiral}, ranging from simple collinear magnetic structures including ferromagnets (FMs) \cite{Spin_Waves_Dyson} and antiferromagnets (AFMs) \cite{AFM_Neel} to non-collinear structures like skyrmion lattices \cite{SkyrmionsBogdanov,SpinWaves_Skyrmions, Topo_Magnons_Diaz_1, Topo_Magnons_Diaz_2}. These magnetically ordered materials host collective excitations of coupled magnetic moments. These excitations can be described classically using spin waves, but are interpreted as particles - known as magnons - when treated quantum mechanically \cite{Spin_Waves_Dyson,Spin_Waves_Kittel, Spin_Waves_NoltingRamakanth}. 
        
The interest in magnons rose in recent years \cite{Report_Magnonics, Magnonic_Roadmap} as they provide a platform for easy-to-manipulate long-distance transport with low dissipation \cite{SpinTransportFMI, SpinTransportAFMI}. They may provide an alternative to the Joule-heating-plagued electrons in information transport and processing \cite{MagnonSpintronics}, as a key computational device, transistors, can be realized by magnetic structures \cite{Magnon_Transistor}.  Moreover, it is possible to link magnetic materials to conventional electronic devices using the (inverse-)spin-Hall effect \cite{Inverse_Spin_Hall_Cornelissen, SpinTransportAFMI}, enabling the integration of magnonic devices with state of the art technology. This is highly significant due to the reduction in computational technology size and the emergence of two-dimensional materials\cite{2DMat, 2DMat2, 2DMatMoS, 2DMatMoSTrans, 2DMatBN, Graphen, GraphenEncBN}, notably Van der Waals (VdW) materials\cite{VdWHeterostructures1, VdWHeterostructures2, VdWHeterostructures3, VdWHeterostructures4}. These materials provide the potential to produce engineered synthetic systems with desired attributes by layering different materials together. Two-dimensional (2D) magnetic layers\cite{2DMagnetism1, 2DMagnetism2, 2DMagnetism3, 2DMagnetism4, 2DMagnetism5, 2DMagnetism6, 2DMagnetism7, 2DMagnetism8, 2DMagnetism9, 2DMagnetism10} represent just one of many potential building blocks. Therefore, it is necessary to enhance our understanding of the excitations of 2D magnetic materials, along with the impact of coupling between distinct magnetic materials, considering the potential for squeezing and entanglement within magnetic systems.
        
    Similar to photons \cite{PhotonSqueezing,SqueezedLight_Wu}, quantum squeezing and entanglement for magnons have been predicted theoretically  \cite{Squeezing_as_Niche,AFM_Squeezing_Akash, AFM_Squeezing_Wuhrer, Non_Integer_Spin_Akash} and observed in experiments \cite{Magnon_Squ_Exp_1, Magnon_Sq_Exp_2}. However, in contrast to photons,  squeezing is an inherent property of magnetic materials. Squeezing for magnons refers to quadrature squeezing. This means decreasing the variance of one observable while increasing the variance of a canonically conjugate observable, such as the spin components in the $x$ and $y$-direction in a squeezed FM with the $z$-axis as easy axis. This is because squeezed magnon states are a superposition of non-squeezed magnon Fock-states. Consequently, the system already holds a substantial number of non-squeezed magnons even in the ground state. 
    This  superposition has been proposed for exciting multiple quantum dots simultaneously and in the process entangle them \cite{Multiple_QDot_Excitation_Skogvoll} or being probed by quantum dots adjacent to the magnet\cite{MagnonOcc_Qubit}. It was also discussed that the entanglement due to magnons in VdW materials is switchable by electronic and magnetic means leading to electrical controllable entanglement of distant qubits\cite{VdW_Entanglement}. As a result, magnons make an interesting tool for quantum computing \cite{Quantum_Internet, Teleport_Squeezed, Quantum_Cryptography1, Quantum_Cryptography2}. 
    
    Our work explores the feasibility of long-distance magnon entanglement through dipole-dipole interaction in two separated FMs, facilitating the entanglement of remote systems or by magnetic materials involved in the formation of VdW materials. By 'long range,' we denote a distance greater then  decay length of the Heisenberg exchange, which experiences an exponential decay.
	
		\begin{figure*}
        \includegraphics[width=2\columnwidth]{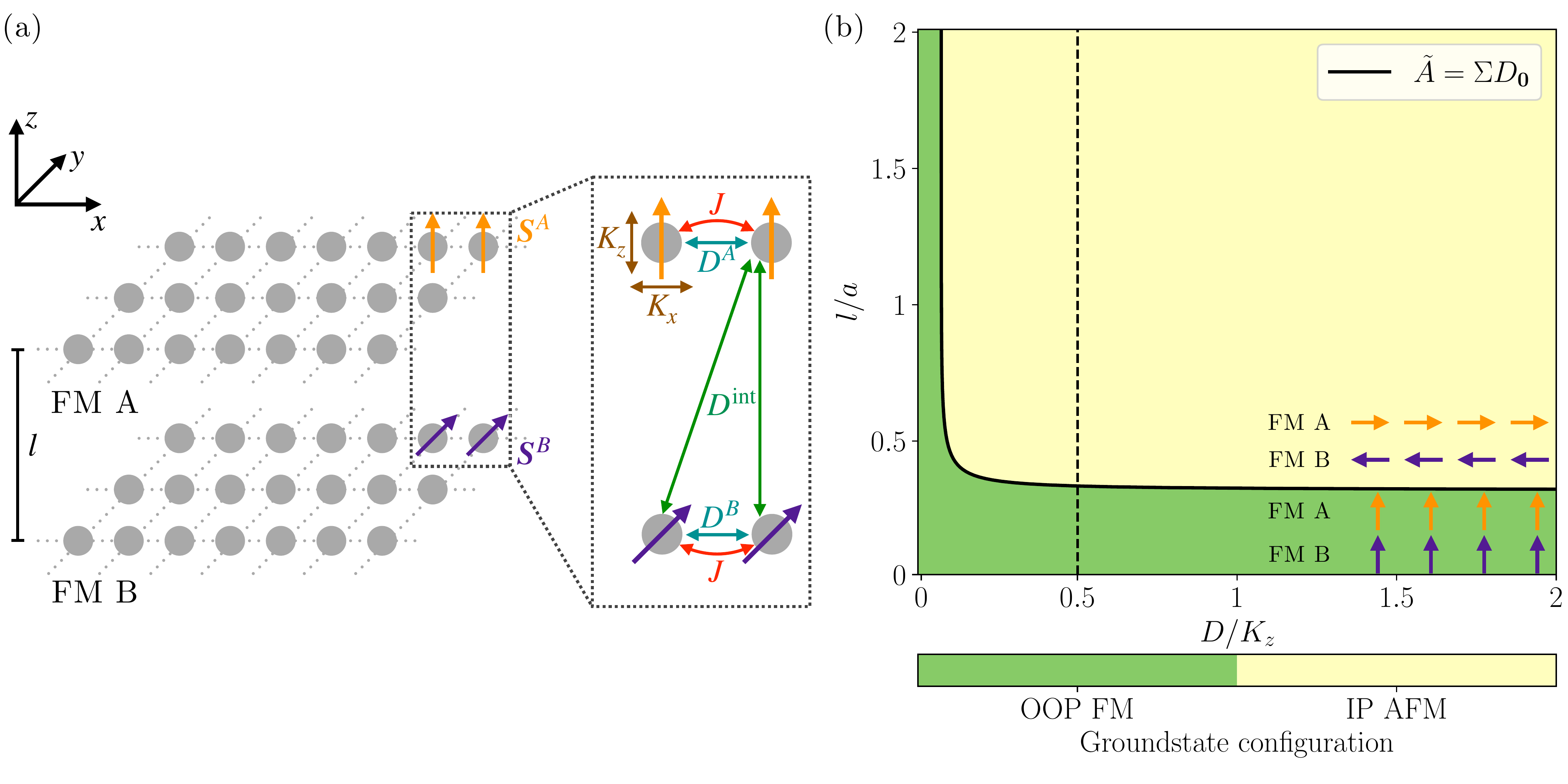}%
		\caption{(a) System of two FMs parallel to the $x$-$y$-plane separated by a distance $l$ along the $z$-axis. The magnified area depicts the interactions between the spins of the different FMs.
		(b) Phase diagram showing the orientation of the spins in FM $A$ and $B$ depending on the distance $l/a$ and the dipole-dipole interaction strength $D/K_z$. For small distances or small dipole-dipole interactions spins in both FM point out of plane forming a ferromagnetic ordering between the FMs (green, OOP FM). For large distances and large dipole-dipole interactions spins in both FM lie in-plane forming an antiferromagnetic ordering between the FMs (yellow, IP AFM). The border between both phases (black solid line) is given by Eq.~\eqref{eq:PhasseBoundary}. The black dashed line shows the phases depending on the distance for $D = 0.5 K_z$. The orientation of the spins inside each FM is depicted by the arrows with the figure-plane  being the $x$-$z$-plane.}
		\label{fig:1}
	\end{figure*} 
	
	As a model system we consider two two-dimensional (2D) square lattice FMs separated by a distance $l$, see Fig.~\ref{fig:1}(a). The spins inside each FM are subjected to ferromagnetic Heisenberg interaction and two uniaxial anisotropies: one perpendicular to the plane of the 2D FMs ($z$-direction) and the other along one of  principle axis of the FMs ($x$-direction). Moreover, we consider the long range interaction between the spins to be transmitted by dipole-dipole interaction.  The Hamiltonian describing the interactions of spins inside FM $A$ is given by 
    \begin{align}
		\oH_{A} &=  \oH_{\text{ex}}^{A} + \oH_{\text{an}}^{A}  + \oH_{\text{D}}^{A}\, ,
		\label{eq:Def_Hamiltonian}
	\end{align}
	with a similar expression for FM $B$. The Heisenberg exchange is given by 
	\begin{align}
		\oH_{\text{ex}}^{A} &= -\sum_{ \substack{\vec r_i,\vec r_j\in {A}\\\vec r_i \neq \vec r_j}} J_{ij}  \vec{\hat S}_i \cdot \vec{\hat S}_j \, ,\label{eq:symm_exch}
	\end{align}
	with $J_{ij} = J_{ji} >0$ being the ferromagnetic exchange strength between spins $\ovS_{i}$ and $\ovS_{j}$ of spin length $S$  at sites $\vec r_i$ and $\vec r_j$ in FM $A$ preferring a parallel configuration of spins.
	
	The anisotropy term in the  Hamiltonian, Eq.~\eqref{eq:Def_Hamiltonian}, is given by 
	\begin{align}
		\oH_{\text{an}}^{A} &=- K_z\sum_{\vec r_i\in A} \left(\hat{S}^z_i\right)^2 -  K_x \sum_{\vec r_i\in A} \left(\hat{S}^x_i\right)^2\, ,
	\end{align}
	where $K_{z}$  ($K_x$) is the strength of the uniaxial anisotropy along the $z$-axis ($x$-axis). We assume $K_z\gg K_x >0 $ such that the spins prefer to align parallel to the $z$-direction. 
	
	Finally, we consider dipole-dipole interaction between magnetic moments given by 
	\begin{align}
		\oH_{\text{D}}^{A} =  \sum_{ \substack{\vec r_i,\vec r_j\in {A}\\\vec r_i \neq \vec r_j}}D_{ij} \left[ \vec {\hat S}_i \cdot  \vec {\hat S}_j - 3 \left(\vec e_{ij}\cdot  \vec {\hat S}_i\right)\left(\vec e_{ij}\cdot \vec {\hat S}_j\right)\right]\, . \label{eq:Def_Dipol}
	\end{align}
	Here $\vec e_{ij}$ is the unit vector along the connection line of spins at $\vec r_i$ and $\vec r_j$.	$D_{ij}$ is the strength of dipolar coupling between two spins at $\vec r_i$ and $\vec r_j$ which is of the form 
	\begin{align}
		D^{A}_{ij} = \frac{a^3 D^{A}}{\left|\vec r_i - \vec r_j\right|^3} = \frac{\mu_0 g^{A}_1 g^{A}_2\mu_B^2}{\hbar ^2} \frac 1 {|\vec r_i - \vec r_j|^3} \, ,\label{eq:Def_Dipol_coeff}
	\end{align} 
	with $\mu_0$ being the vacuum magnetic permeability, $g_i^A$ the gyromagnetic ratio of the interacting magnetic moments, $\mu_B$ the Bohr magneton, $\hbar$ the reduced plank constant and $a$ the lattice constant. Throughout this work, we will assume that $\hbar, a = 1$, which means that all the interaction parameters ($J_{ij}$, $K_x$, $K_z$, $D^{A}$) will be given in the unit of energy.  
 
	The dipole-dipole interaction decays with the inverse of the third power of the distance between the spins and thus is classified as long range.

    The coupling between both FMs is also conveyed through dipolar interaction
  	\begin{align}
  		\oH_{\text{int}} &=\!   \sum_{\substack{\vec r_i,\in A\\\vec r_j\in B}}  D^{\text{int}}_{ij} \left[\vec {\hat S}_i \cdot \vec{\hat S}_j - 3 \left(\!\vec e_{ij}\cdot \vec{\hat S}_i\!\right)\left(\!\vec e_{ij}\cdot \vec {\hat S}_j\!\right)\right]\, .\label{eq:Int_Ham}
  	\end{align}
   	We assume similar FMs. Thus, Eq.~\eqref{eq:Def_Dipol_coeff} yields similar dipole constants for the interaction inside each FM and between both FMs and we assume $D^{A/B} = D^{\text{int}} = D$.  

    The linear spin wave theory explains the excitation within magnetic systems as harmonic excitations, spin waves, by expanding the spin operators around the classical ground state using magnonic creation and annihilation operators. The classical configuration for each isolated FM is ferromagnetic, because we assume that the exchange interaction is dominant inside each FM. The anisotropies restrict the direction of magnetization to lie within the $x$-$z$-plane. The exact orientation is determined by the strength of the dipole-dipole interaction inside each and between both FMs. Thus the classical ground state depends on the dipole interaction strength $D$ and the distance $l$ between both FMs. Only two orientations are taken by the FMs, see Fig.~\ref{fig:1}(b).

    For short distances a ferromagnetic out-of-plane configuration (OOP FM) between both FMs is taken due to the dominance of the dipole interaction between the systems, which favours a parallel, ferromagnetic alignment along the connecting line ($\vec e_z$). For large separations and weak dipole-dipole interaction also the OOP FM configuration is taken. This is due to the fact that with $D$ not only the interaction between the FMs is varied, but also the dipole-dipole interaction between spins inside each FM. Therefore, for weak dipole-dipole interaction the spins prefer to align with $K_z$.
  	
	For significant distances, the spins align ferromagnetically in-plane due to the dominance of the dipole-dipole interaction within each FM over the anisotropy.  Minimising the interaction term when the spins are perpendicular to the connection line results in an AFM ordering between the two ferromagnets. This is referred to as in-plane AFM (IP AFM) ordering.
  	
  	In this work we investigate square lattice FMs, where the phase boundary between OOP FM and IP AFM phase, seen as black solid line in Fig.~\ref{fig:1}(b), is given by
  	\begin{align}
  		 K_z - (K_x + 3D_{x|\vec 0}) &=  D_{\vec 0}^{\text{int}} - \frac 32\left(D_{x|\vec 0}^{\text{int}} + D_{z|\vec 0}^{\text{int}}\right)\, .\label{eq:PhasseBoundary}
  	\end{align}
    The coefficient(s) $D_{\alpha | \vec 0}$ $(D_{\vec 0}^{\mathrm{int}}, D_{\alpha|\vec 0}^{\mathrm{int}})$ is (are) the sum(s) of the dipole interaction inside (between) the FMs, with $\alpha$ specifiying the direction of the connection vector. The exact definition is given in the supplementary part\cite{Supplementary}.
    \begin{figure}
  		\begin{center}
  			\includegraphics[width = \columnwidth]{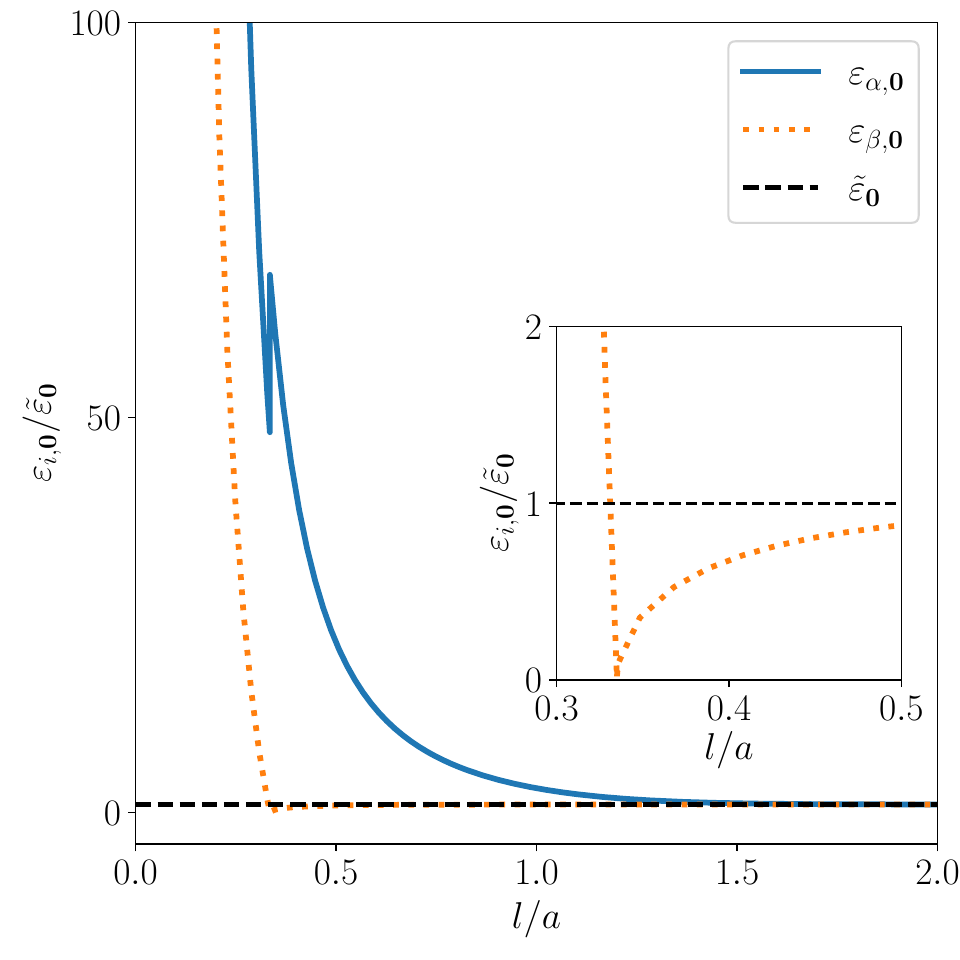}
  		\end{center}
  		\caption{Energy of the $\vec k = \vec 0$ squeezed magnon modes depending on the distance $l/a$ between both FMs in units of the energy of the eigenmodes in non-interacting FMs, $\tilde \varepsilon_{\vec 0}$. At the phase change, $l/a \approx 0.33$,  the energy $\varepsilon_{\beta,\vec 0}$ vanishes. For large distances $\varepsilon_{\alpha,\vec 0}$ and $\varepsilon_{\beta,\vec 0}$ approach $\tilde{\varepsilon}_{\vec 0}$ (black, dashed line). Parameters are chosen as $D = 0.5\cdot 10^{-4} |J_1|$, $K_z = 10^{-4}|J_1|$ and $K_x = 10^{-6}|J_1|$.}
  		\label{fig:3}
  	\end{figure}
    \begin{figure*}
  		\subfloat[\label{fig:4a}]{%
  			\includegraphics[width=\columnwidth]{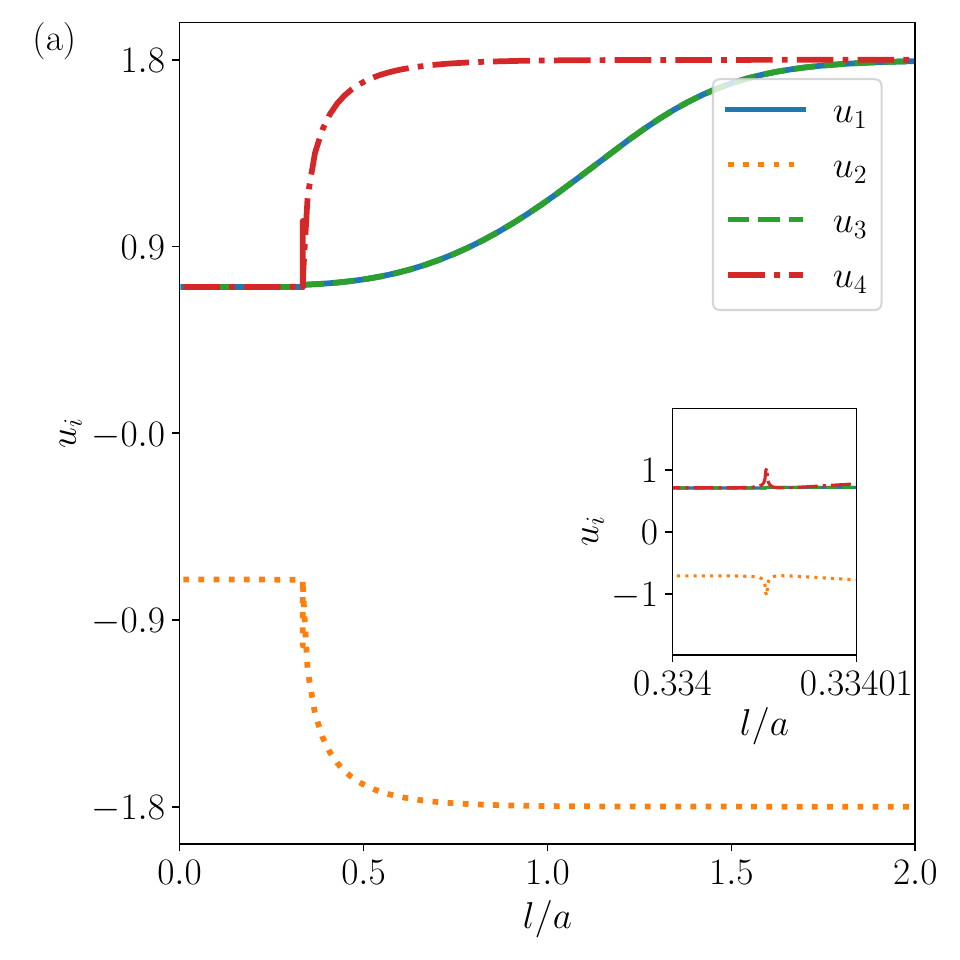}%
  		}\hspace*{\fill}%
  		\subfloat[\label{fig:4b}]{%
  			\includegraphics[width=\columnwidth]{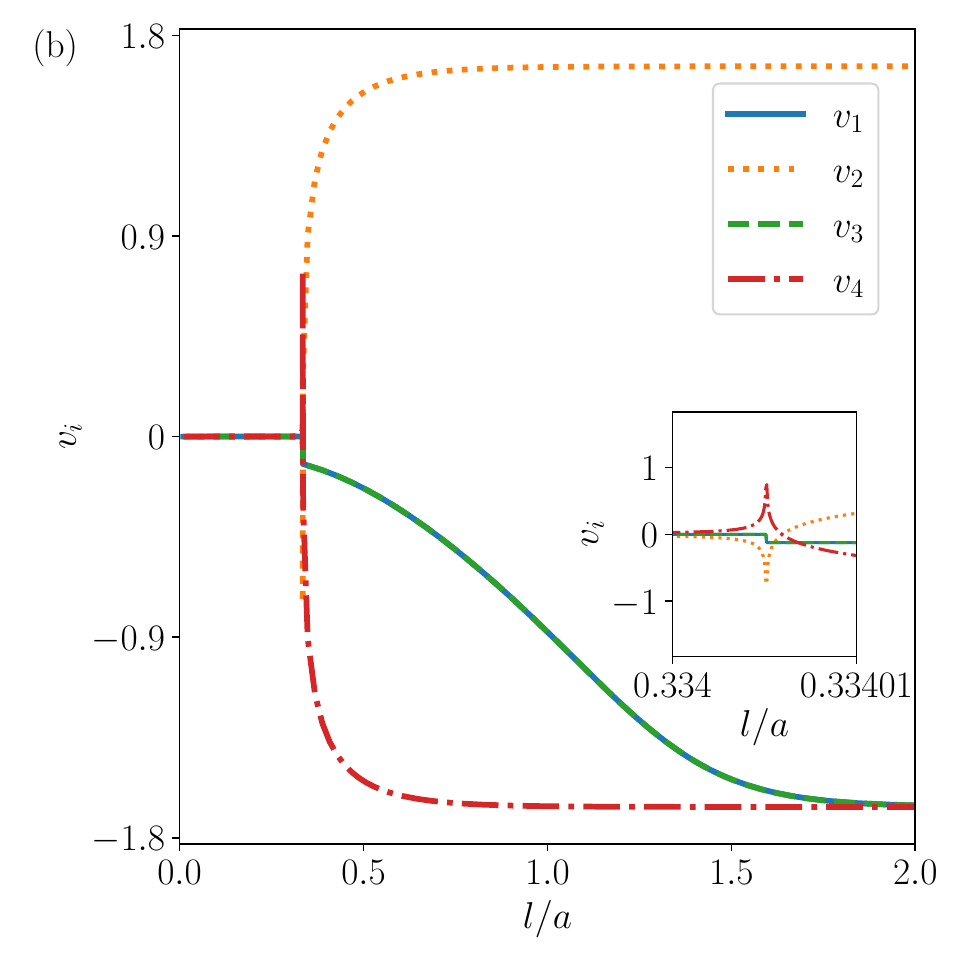}%
  		}
  		\caption{Dependence of the  Bogoliubov matrix elements on the distance $l$ for the $\vec k = \vec 0$ mode with $D = 0.5\cdot 10^{-4} |J_1|$, $K_z = 10^{-4}|J_1|$ and $K_x = 10^{-6}|J_1|$. Divergences and jumps occur at the phase change $l/a \approx 0.33 $
  			(a) Elements of $U_{\vec 0}$. 
  			(b) Elements of $V_{\vec 0}$.}
  		\label{fig:4}
  	\end{figure*} 

   \noindent The classical configuration is represented by the right handed eigensystems 
	\begin{align}
  	 	\vec e_{i,1} = \begin{pmatrix}
  	 		\cos(\vartheta_i)\\0\\-\sin(\vartheta_i)
  	 	\end{pmatrix}\, &, \qquad \vec e_{i,2} = \begin{pmatrix}
  	 		0\\1\\0
  	 	\end{pmatrix} \, \notag\\ 
  	 	\vec e_{i,3} =& \begin{pmatrix}
  	 		\sin(\vartheta_i)\\0\\\cos(\vartheta_i)\, ,
  	 	\end{pmatrix}\label{eq:Eigensystem}
  	 \end{align}
  	 for $i\in \{A,B\}$, where $\vartheta_A$ ($\vartheta_B$) is the angle between the classical spin configuration of FM $A$ ($B$) and the $z$-direction. Applying the linear Holstein-Primakoff transformation\cite{Holstein_Primakoff_Trafo} yields for spins in FM $A$
    \begin{gather}
        \oS_{i,1} \approx\hbar  \sqrt{\frac{S}{2}}\left(\oa_i + \oad_i \right)\,,  \qquad \oS_{i,2} \approx -i\hbar  \sqrt{\frac{S}{2}}\left(\oa_i -\oad_i \right) \, ,\notag\\ \oS_{i,3} \approx \hbar \left(S - \ona{i}\right)\, ,\label{eq:lin_spin_wave_def_A}
    \end{gather}
    and in FM $B$
    \begin{gather}
        \oS_{i,1} \approx \hbar \sqrt{\frac{S}{2}}\left(\ob_i + \obd_i \right)\,,  \qquad \oS_{i,2} \approx -i\hbar  \sqrt{\frac{S}{2}}\left(\ob_i -\obd_i \right) \, ,\notag\\ \oS_{i,3} \approx\hbar \left(S - \onb{i}\right)\, ,\label{eq:lin_spin_wave_def_B}
    \end{gather}
	with $	\oS_{i,\alpha} = \vec e_{A/B,\alpha} \cdot \ovS_{i}$.  

    We perform the Fourier transformation to take account for the periodic structure of the system
	\begin{align}
		\oa_{\vec k} &= \frac 1 {\sqrt N} \sum_{\vec r_i\in A} e^{-i \vec r_i \cdot \vec k} \oa_i \, , &  \ob_{\vec k} &= \frac 1 {\sqrt N} \sum_{\vec r_i\in B} e^{-i \vec r_i \cdot \vec k} \ob_i \, ,\label{eq:Fourier}
	\end{align}
	where we assume  $N_A = N_B = N$. Inserting Eqs.~(\ref{eq:lin_spin_wave_def_A}-\ref{eq:Fourier}) into Eqs.~(\ref{eq:Def_Hamiltonian}+\ref{eq:Int_Ham}) the Hamiltonian takes the form 
    \begin{align}
	\oH 
	&= \sum_{\vec k \in \text{BZ}^{+}}  \ova^\dagger_{\vec k} \underbrace{\begin{pmatrix}
	E_{A,\vec k}& \mu_{1,\vec k}^\ast& 2 \xi_{A,\vec k}^\ast & \mu_{2,\vec k}^\ast \\
	\mu_{1,\vec k} &E_{B,\vec k} & \mu_{2,\vec k}^\ast  & 2 \xi_{B,\vec k}^\ast\\
	  2 \xi_{A,\vec k}& \mu_{2,\vec k} & E_{A,\vec k}& \mu_{1,\vec k}\\
	 \mu_{2,\vec k} & 2 \xi_{B,\vec k}& \mu_{1,\vec k}^\ast &E_{B,\vec k} 
	\end{pmatrix}}_{H_{\vec k}}\ova_{\vec k}\, ,
	\label{eq:Hamiltonian_matrix}
	\end{align}
	where we defined $\ova_{\vec k} = \left(	\oa_{\vec k}, \ob_{\vec k} , \oad_{-\vec k} , \obd_{-\vec k}\right)^\top$ and $\text{BZ}^+$ labels all $\vec k$ in the Brillouin zone with a positive (or zero) $k_x$ component. The exact definition of each parameter contained by the Hamiltonian is given in the supplementary\cite{Supplementary}.
	
	The meaning of the different parameters, if only they together with the diagonal elements $E_{i, \vec k}$ would appear in the Hamiltonian,  is outlined below. Energy of a bare magnon mode $\oa_{\vec k}$ ($\ob_{\vec k}$)  inside FM $A$ ($B$) is given by $E_{A,\vec k}$ $\left(E_{B,\vec k}\right)$, while $\mu_{1,\vec k}$ results in a hybridisation of the magnon modes of the different FMs. This does not change the vacuum state and corresponds to a rotation in the phase space spanned by the $A$ and $B$  magnons.
	
	Squeezing inside FM $A$ ($B$) is mediated by $\xi_{A,\vec k}$ ($\xi_{B,\vec k}$) entangling modes with wave vectors $\pm \vec k$ inside each FM, while squeezing of $\pm \vec k$ modes of the different FMs is a result of   $\mu_{2,\vec k}$.  Squeezing alters the vacuum state as the squeezed magnon states are superpositions of populated  $A$ and $B$ magnon states \cite{AFM_Squeezing_Wuhrer}. If we combine different of these parameters a clear distinction between squeezing and hybridization for each parameter cannot be given anymore. 
	 
	 In order to diagonalize the Hamiltonian we perform a four-dimensional (4D) Bogoliubov transformation \cite{Bogo_Trafo, Valatin_Trafo} which is a symplectic transformation converting our bare magnon operators $\oa_{\vec k}$ and $\ob_{\vec k}$ into squeezed magnon operators $\oal_{\vec k}$ and $\obe_{\vec k }$. The Bogoliubov transformation can be written as
	 	\begin{figure} 
	 			\includegraphics[width=\columnwidth]{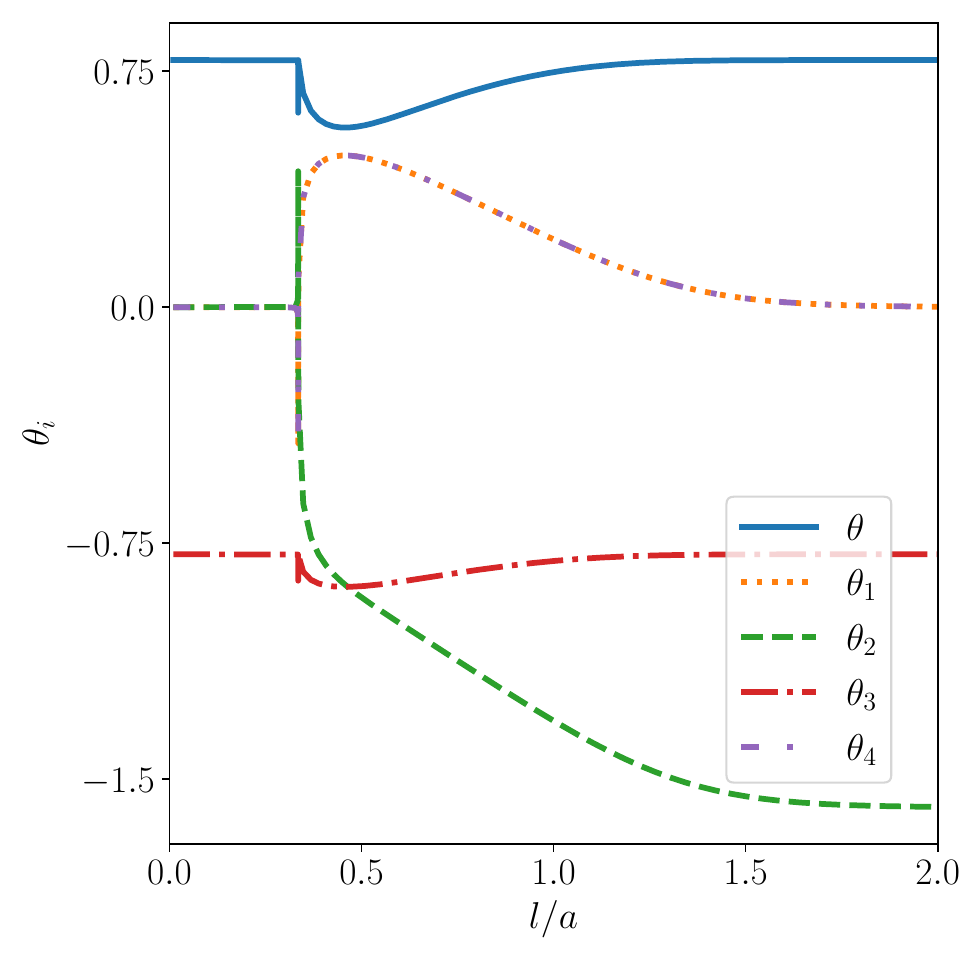}
	 	\caption{Dependence of the squeezing parameters $\theta, \theta_{1}, \theta_{2}, \theta_{3} , \theta_{4}$ on the distance $l$ for the $\vec k = \vec 0$ mode. Divergences occur at the  phase change $l/a \approx 0.33$.
	 		Parameters are $D = 0.5\cdot 10^{-4} |J_1|$, $K_z = 10^{-4}|J_1|$ and $K_x = 10^{-6}|J_1|$.}
	 	\label{fig:5}
	 \end{figure} 
	 \begin{align}\ova_{\vec k} &= 
	 	 \begin{pmatrix}
	 		\oa_{\vec k}\\ \ob_{\vec k} \\ \oad_{-\vec k} \\ \obd_{-\vec k}
	 	\end{pmatrix} = \begin{pmatrix}
		 	U_{\vec k} & V_{\vec k}\\V_{\vec k}^\ast & U_{\vec k}^\ast
	 	\end{pmatrix} \begin{pmatrix}
	 	\oal_{\vec k}\\ \obe_{\vec k} \\ \oald_{-\vec k} \\ \obed_{-\vec k}
 	\end{pmatrix}= \mathcal{U}_{\vec k} \ovaL_{\vec k }\, ,\label{eq:Bogotrafo}
	 \end{align} 
 	 with 
 	 \begin{align}
 	 	U_{\vec k}^{\phantom{\dagger}}U_{\vec k}^\dagger - V_{\vec k}^{\phantom{\dagger}}V_{\vec k}^\dagger &= 1 \, ,& 	U_{\vec k}^{\phantom{\top}}V_{\vec k}^\top - V_{\vec k}^{\phantom{\top}}U_{\vec k}^\top = 0\, .
 	 \end{align}
	 The elements of $U_{\vec k}$ are responsible for hybridization and the elements of $V_{\vec k}$ for squeezing, with
	 \begin{align}
	 	U_{\vec k} &= \begin{pmatrix}
	 		u_{1,\vec k} & u_{2, \vec k}\\ u_{3, \vec k} & u_{4, \vec k}
	 	\end{pmatrix}\, , & V_{\vec k} &= \begin{pmatrix}
	 	v_{1,\vec k} & v_{2, \vec k}\\ v_{3, \vec k} & v_{4, \vec k}
 	\end{pmatrix}\, .
	 \end{align}
 	Solving the eigenvalue equation of the Hamiltonian yields the elements of the Bogoliubov matrix $\mathcal U_{\vec k}$ and the eigenenergies $\varepsilon_{\alpha/\beta,\vec k}$ of the squeezed magnon modes 
 	\begin{align}
 		\varepsilon_{\alpha/\beta,\vec k} &= \sqrt{\left|E_{\vec k} \pm \mu_{1,\vec k}\right|^2 - \left|\mu_{2,\vec k} \pm 2 \xi_{\vec k}\right|^2}\, ,
 	\end{align} 
 	where we used that we cope with similar FMs yielding $E_{A,\vec k} = E_{B,\vec k} = E_{\vec k}$ and $ \xi_{A,\vec k} = \xi_{B,\vec k} = \xi_{\vec k}$ and  that we regard  either the OOP FM or the IP AFM phase. Fig.~\ref{fig:3} shows the energies of magnons $\oal_{\vec 0}$ and $\obe_{\vec 0}$ depending on the distance. We see that for large distances both converge towards the energy $\tilde \varepsilon_{\vec 0} = \sqrt{\varepsilon_{\vec 0}^2 - 4 |\xi_{\vec 0}|^2}$ of the squeezed magnons in the case of non interacting FMs. At the phase change $\varepsilon_{\alpha,\vec 0}$ shows a jump, while  $\varepsilon_{\beta,\vec 0}$ vanishes.
 	
From several two-mode squeezed cases we expect the largest squeezing and entanglement for the uniform $\vec k = \vec 0$ mode \cite{AFM_Squeezing_Wuhrer, Spin_spiral_Wuhrer}. Thus, in the following, we will limit our considerations to the $\vec k = 0$ mode and will drop the wave vector index for clarity. For the uniform mode all parameters in Eq.~\eqref{eq:Hamiltonian_matrix} are real yielding real Bogoliubov parameters of the form
\begin{align}
 	u_{1} &=\! \sqrt{\!\frac{E \!+\! \mu_{1} \!+\! \varepsilon_{\alpha}}{4 \varepsilon_{\alpha}}} , & u_{2} &=\! - \sqrt{\!\frac{E\! - \!\mu_{1}\! - \!\varepsilon_{\beta}}{4 \varepsilon_{\beta}}}\, ,
\end{align}
with the remaining matrix elements given in the supplementary material\cite{Supplementary}. Fig.~\ref{fig:4} shows the dependence of the Bogoliubov matrix elements on the distance of both FMs. For distances  $l/a < 0.33$, in the OOP FM phase, we have non-zero values for all elements of $U$ and close to zero value for all elements of $V$. Thus, the ferromagnetic structure is dominated by hybridisation ($U$) with little to no squeezing ($V$).  Due to the vanishing magnon energy of the  $\obe$ magnons at the phase change $l/a \approx 0.33$, we see divergences in $u_{2}$, $u_{4}$, $v_{2}$ and $v_{4}$. 

In the IP AFM configuration ($l/a > 0.33$) we see an increase in all elements of  $\mathcal{U}$ compared to the OOP FM configuration with the elements responsible for squeezing  ($V$) being comparable in magnitude.  This is due to the large inherent degree of squeezing of AFMs \cite{AFM_Squeezing_Wuhrer}. 	

The Bogoliubov transformation for the system at hand is an element of the ten dimensional group of 4D symplectic matrices  $\mathrm{Sp}(4,\mathds R)$ \cite{Symplectic_Group}.  The limitation to the $\vec k = 0$ mode leaves us with real a Bogoliubov matrix only which can be parametrised by four parameters.  We connect the Bogoliubov matrix to the generators of the group and introduce the four-mode squeezing operator
 	\begin{align}
 		\hat S_4(\vec \theta) &= \exp{\left(\sum_{i = 1}^{4} \theta_{i} \hat \Phi_{i}\right)}\, ,\label{eq:SqueezingOperator}
 	\end{align}
 	with $\theta_i$ being the real parameter corresponding to the generator $\hat \Phi_{i}$ which in  quantum representation are  given by 
 	\begin{align}
 		\hat \Phi_{1/2} &=  \oad\oad - \oa\oa\mp \left(\obd\obd- \ob\ob\right)\, ,\\
 		\hat \Phi_{3} &= \oad\ob- \oa \obd  + \oad\ob - \oa\obd\, ,\\
 		\hat \Phi_{4} &= \oad\obd- \oa\ob + \oad\obd- \oa\ob\, .
 	\end{align}
 	Each of these (pair of) generators  can  be identify as a one or two-mode operation. $\theta_{1} \hat \Phi_{1}+\theta_{2}\hat \Phi_{2}$ results in one-mode squeezing of  modes inside FM $A$  and $B$ with squeezing parameters $r^{A/B}=\theta_{2} \pm \theta_{1}$ for FM $A$ ($B$). $\hat \Phi_{3}$ ($\hat \Phi_{4}$) results in hybridization (two-mode squeezing) of the  magnon modes  of the different FMs. Thus, isolated each of the parameters $\theta_{i}$ can be identified with either one-mode, two-mode squeezing or hybridisation. However, this clear separation cannot be made if  multiple parameters are involved,  as the different generators do not commute. Ongoing we will refer to all $\theta_{i}$  as "squeezing parameters".

 	We use the properties of the symplectic group to connect the elements of the Bogoliubov matrix to the squeezing parameters: 
	\begin{align}
		\ovaL = \mathcal U^{-1}\ova = \hat S_4(\vec \theta)\ova\hat S_4^{-1}(\vec \theta)\, ,\label{eq:Connection_Bogo_Sq}
	\end{align} 	
	which yields
	\begin{align}
		u_{1} &=  \cosh(\theta_{2}) \cos(\theta)  + \frac{\theta_{1}}{\theta}\sin{(\theta)} \sinh{(\theta_{2})}\, ,\\
		u_{2} &=  \frac{\theta_{4}}{\theta}\sin{(\theta)} \sinh{(\theta_{2})} + \frac{\theta_{3}}{\theta}\sin{(\theta)} \cosh{(\theta_{2})}\,,
	\end{align}
	with $\theta= \sqrt{(\theta_{3})^2 - (\theta_{1})^2 -(\theta_{4})^2}$.  The remaining matrix elements expressed in terms of the squeezing parameters are given in the supplementary material\cite{Supplementary}. 
	
	Fig.~\ref{fig:5} shows the dependence of the squeezing parameters on the distance. They show a similar behaviour  as the elements of $\mathcal U$. The parameters identified with intra FM ($\theta_1, \theta_2$) and inter FM ($\theta_4$) squeezing are close to zero in the OOP FM phase and finite in the IP AFM phase, while the parameter identified with inter FM hybridization ($\theta_5$) is of a finite value for the whole parameter range.  This behaviour supports our identification based on the generators $\Phi_i$. The divergence of the squeezing parameters is a result of vanishing energy of the $\obe_{\vec  0}$ magnon modes at the phase change. 
	
	To determine the entanglement between the $\vec k = 0$ magnon modes $\oa$ and $\ob$  in the squeezed vacuum state 
	\begin{align}
		\left|0\right>_{\text{Sq}} = \hat S_4(\vec \theta)\left(\left|0\right>_{A}\otimes\left|0\right>_{B}\right)\, ,
	\end{align}
	of the system, we use the logarithmic negativity \cite{Log_negativity}
	\begin{align}
		E_N &=\max{\{ 0,- \ln{\left(2\eta^-\right)}\}}\, ,\label{eq:Entanglement}
	\end{align}
	as entanglement measure, with $\eta^-$ being the lowest symplectic eigenvalue of the symmetric covariance matrix $V$ given by 
	\begin{align}
		V_{ij} &= \frac 1 2 \left<\left\{\hat R_i, \hat R_j\right\}\right> \, .\label{eq:CovMat}
	\end{align}
	The anticommutator is represented by $\left\{.,.\right\}$ and $\vec{\hat R} $ is the phase space vector with 
	\begin{align}
	\vec{\hat R} &=  	\begin{pmatrix}
			\hat q_A\\\hat q_B\\\hat p_A\\\hat p_B
		\end{pmatrix} =\underbrace{ \frac 1 {\sqrt 2} \begin{pmatrix}
		\mathds 1_{2\times 2} &  \mathds 1_{2\times 2}\\
		-i\mathds 1_{2\times 2} & i \mathds 1_{2\times 2}
	\end{pmatrix}}_{M}\ova =  M \mathcal U \ovaL\, ,\label{eq:PhaseSpace}
	\end{align}
	where we used Eq.~\eqref{eq:Bogotrafo}. The lowest symplectic eigenvalue in terms of the squeezing parameters is given by
	\begin{widetext}
		\begin{align}
			\eta^- = \frac 1 2& \left|2 \left|\frac{\theta_{4}}{2 \theta} \sin{\left(2\theta\right)} -  \frac{\theta_{3}\theta_{1}}{\theta^2} \sin^2{\left(\theta\right)}\right|-\sqrt{1+ 4\left|\frac{\theta_{4}}{2 \theta}  \sin{\left(2\theta\right)}-  \frac{\theta_{3}\theta_{1}}{\theta^2} \sin^2{\left(\theta\right)}\right|^2 }\right|\, ,\label{eq:eta}
		\end{align}
	\end{widetext}
	which is the central result of this work. 

    	\begin{figure}
		\begin{center}
			\includegraphics[width = \columnwidth]{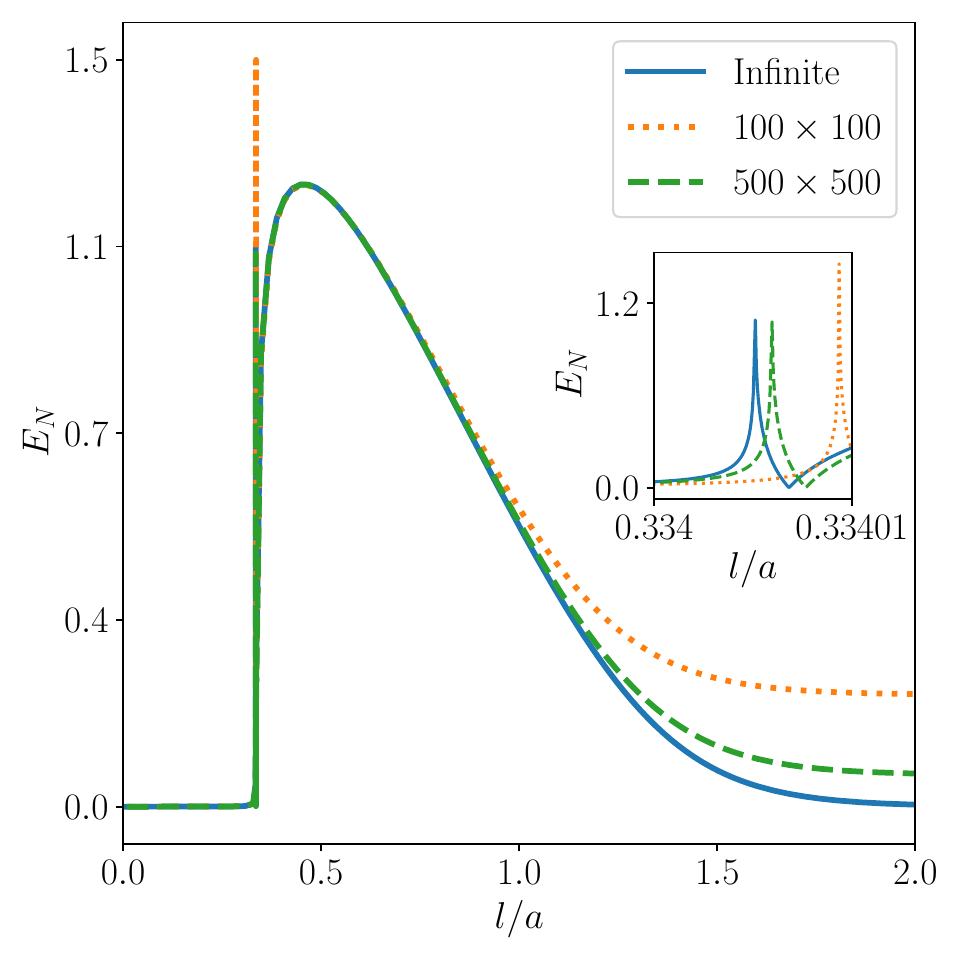}
		\end{center}
		\caption{Entanglement of the $\vec k = \vec 0$ magnon modes in FM $A$ and FM $B$ for infinite (blue, solid), $100\times 100$ (orange, dotted) and $500\times 500$ (green, dashed) sites FMs. We used $D = 0.5\cdot 10^{-4} |J_1|$, $K_z = 10^{-4}|J_1|$ and $K_x = 10^{-6}|J_1|$.}
		\label{fig:6}
	\end{figure} 
 
	From Eq.~\eqref{eq:eta} we see that all squeezing parameter influence the entanglement except for $\theta_2$, because $\Phi_2$ commutes with all other generators. To further investigate the entanglement we concentrate on limiting cases. If only squeezing inside each FM is present ($\theta_{3}, \theta_{4} = 0, \, \theta= i |\theta_{1}|$)  entanglement between modes of the different FM vanishes, which is expected as there is no interaction between the FMs. 
	
	Only hybridisation between the FMs  ($\theta_{1}, \theta_{2},  \theta_{4}  = 0, \, \theta =  |\theta_{3}|$) results also in a vanishing entanglement as the vacuum for hybridised modes is equal to the vacuum of the bare modes $\oa$ and $\ob$ and thus is a separable state.
	
	Two-mode squeezing between both FMs  ($\theta_{1}, \theta_{2},  \theta_{3}  = 0, \, \theta = i |\theta_{4}|$) results in entanglement linear in the squeezing parameter $E_N = 2 |\theta_{4}|$ similar to the entanglement in case of simple two-mode squeezing \cite{Report_Magnonics}. Therefore, known results are reproduced by the analytic formula given in Eq.~\eqref{eq:eta}.  
    
    The distance dependence of the entanglement can be seen in Fig.~\ref{fig:6} where we regard finite and infinite 2D FMs.Fig.~\ref{fig:6} shows similar behaviour of the entanglement for all lattice sizes for $l/a \leq1$.  The phase transition from an OOP FM  configuration to an IP AFM configuration appears around $l/a \approx 0.33$,  in the case of the infinite lattice, and entails a divergence of the entanglement. This is shifted to larger distances for finite lattices. The divergence once again arises due to the vanishing magnon energy at the phase change.
	
	Aside from the divergence the entanglement takes it maximum value at approximately $l/a \approx 0.5$, rapidly decreases afterwards and,  in the infinite case, vanishes already at $l/a \approx 2$. The origin of the maximum is still up to debate and subject to future work. The finite cases with $100\times100$ and $500\times500$  lattices do show a plateau in the entanglement around $l/a \approx 2$, which is higher the smaller the lattice is. Increasing the distance up to the order of the system diameters ($100 a$ or $500a$ respectively) results in a decrease of entanglement that is proportional to $1/l^3$. This behaviour is derived from the source of interaction: the dipole-dipole interaction, Eq.~\eqref{eq:Int_Ham}.
 
	We infer that no long range entanglement between infinite large magnets is present due to dipole interaction, while  for finite magnets the entanglement can take a finite value. Thus, for large magnetic structures dipole interaction alone will not be enough to establish a long range entanglement between magnetic materials and other interactions or structures need to be considered.  We remark that we always assumed periodic boundary conditions for magnons in finite systems. Also, for small distances ($l/a < 1$), other interactions between the magnetic moments may be of significance which require investigation in future work.

    At this point we would like to emphasise that we have studied an experimentally easily realizable system of two distant FMs. While the exact structure may differ from a simple cubic lattice structure, the main point, the interaction between the FMs via the dipole-dipole interaction, is present in any magnetic structure. The importance of this work is further emphasized by the possibility of using the entangled magnons in finite size magnets to establish long-range entanglement between spin-qubits, as proposed by Skogvoll et al. \cite{Multiple_QDot_Excitation_Skogvoll} or Yuan et al. \cite{VdW_Entanglement} This is important for the future experimental realization of quantum computing \cite{Quantum_Internet, Teleport_Squeezed, Quantum_Cryptography1, Quantum_Cryptography2}.
 
	We investigated the entanglement of magnon modes in two 2D ferromagnets coupled by dipole-dipole interaction by representing  the Bogoliubov matrix as an element of the symplectic group to represent its matrix elements in terms of four squeezing parameters, each identified with either two mode squeezing or hybridisation. 
	
	Investing the Bogoliubov matrix elements and the squeezing parameters shows a clear dominance of hybridization over squeezing in the OOP FM phase and a significant contribution of squeezing in the IP AFM phase, which is in agreement with already known results. The system shows a finite hybridization for all distances due to the long range character of the dipole-dipole interaction, while the squeezing vanishes for large distances. 
	
	Using the logarithmic negativity, we derived an analytic expression for the entanglement in terms of the  squeezing parameters and were able to reproduce  known limiting cases and showed that besides squeezing between both FMs also squeezing inside one FM paired with hybridization between both FMs is enough to ensure entanglement between both systems,  while hybridization alone is not enough. This clearly shows the importance of squeezing for the entanglement of magnon modes. For infinite systems, the entanglement already becomes insignificant at $l/a =2$ implying no long range entanglement. However, for finite systems we see a finite plateau in the entanglement before it tends towards zero for distances of the order of the system diameter. The value of this plateau decreases with increasing system size. 

     Concerning potential applications we note that the current high interest in 2D van-der-Waals magnets \cite{2DMagnetism1} are explored experimentally worldwide. These systems are usually fabricated in small flakes and our prediction might be relevant to discuss the entanglement of two (finite-size) flakes of two-dimensional ferromagnetic materials separated by an insulating layer, thus avoiding a direct electronic contact. These entanglement properties may have applications in quantum computing\cite{Multiple_QDot_Excitation_Skogvoll}. Future investigations could concentrate on the possible tuning of entanglement between two close FM structures by an applied magnetic field, which changes the configuration taken by the system.

    \section*{Supplementary}
    \noindent We offer supplemental material\cite{Supplementary} in which extended calculations regarding the classical ground state, quantum ground state and squeezing parameters are given. 
    
    \begin{acknowledgments}
        \noindent This work was financially supported by the Deutsche Forschungsgemeinschaft (DFG, German Research Foundation) via the Collaborative Research Center SFB 1432 project no.~425217212 and project no. 417034116.
    \end{acknowledgments}

    \section*{Authors declaration}
    \noindent The authors have no conflicts to disclose. 

    \section*{Data Availability Statement}
    \noindent The data that support the findings of this study are available from the corresponding author upon reasonable request.

    \section*{Authors contributions}
    \noindent\textbf{Dennis Wuhrer:} Formal analysis (lead); software (lead); writing -- original draft (lead); vizualization (lead); writing -- review and editing (equal); methodology (equal); 
    \textbf{Niklas Rohling:} Formal analysis (supporting); software (supporting); supervision (supporting);  writing -- original draft (supporting); writing -- review and editing (equal). 
    \textbf{Wolfgang Belzig:} Conceptualization (lead); Funding acquisition (lead); methodology (equal); project administration (lead); Supervision (lead); writing -- original draft (supporting) writing -- review and editing (equal).

\bibliography{bib}

\end{document}